\documentclass[twocolumn]{aa}
\usepackage{graphicx}
\usepackage{txfonts}
\begin{document}
\title{Using globular clusters to test gravity in the weak acceleration regime}

   \subtitle{}

   \author{R. Scarpa\inst{1}
          G. Marconi\inst{1}\and R. Gilmozzi\inst{1}\fnmsep 
          }

   \offprints{R. Scarpa; rscarpa@eso.org}

\institute{European Southern Observatory, 3107 Alonso de Cordova, 
Santiago, Chile}

   \date{}
\authorrunning{Scarpa, Marconi \& Gilmozzi}
\titlerunning{Testing gravity in the weak acceleration regime}

\abstract{ We have carried out a study of the velocity dispersion of
the stars in the outskirts of the globular cluster $\omega$ Cen, finding
that the velocity dispersion remains constant at large radii rather
than decrease monotonically. The dispersion starts to be constant for
an acceleration of gravity of $a=2.1^{+0.5}_{-0.4}\times 10^{-8}$ cm
s$^{-2}$.  A similar result is obtained reanalyzing existing data for
the globular cluster M15 where the profile flattens out for
$a=1.7^{+0.7}_{-0.5}\times 10^{-8}$ cm s$^{-2}$.  In both cases the
acceleration is comparable to that at which the effect of dark matter
becomes relevant in galaxies.  Explanations for this result within
Newtonian dynamics exist (e.g. cluster evaporation, tidal effects,
presence of dark matter) but require ad hoc tuning of the relevant
parameters in order to make in both clusters the dispersion profile
flat starting exactly at the same acceleration. We suggest that this
result, together with a similar one for Palomar 13 and the anomalous
behavior of spacecrafts outside the solar system, may indicate a
breakdown of Newton's law in the weak acceleration regime. Although
not conclusive, these data prompt for the accurate determination of
the internal dynamics of as many GCs as possible.   }

\maketitle
%

\section{Introduction}

Newton's law of gravity is routinely used to describe the physical
properties of galaxies, even though its validity has been fully and
directly verified only within the boundaries of the solar system.  At
the distance of Pluto, the sun produces an acceleration of $\sim
3\times 10^{-4}$ cm s$^{-2}$. By contrast, the acceleration
experienced by the sun in the gravitational field of the Milky Way is
$2\times 10^{-8}$ cm s$^{-2}$. Thus, any time Newton's law is applied
to galaxies (e.g., to infer the existence of dark matter, hereafter
DM), its validity is extrapolated by several orders of magnitude.
Although there are in principle no reasons to distrust Newton's law in
the weak acceleration regime, testing it would be interesting on its
own.  That such a test may be important is also suggested by the
strong observational evidence that all spacecraft in the outer solar
system and beyond are experiencing an anomalous, unexplained
additional acceleration toward the sun (\cite{anderson98}).

In addition, a particular modification of Newtonian dynamics known as
MOND (\cite{milgrom83}) claims to be able to describe many properties
of galaxies without invoking DM (\cite{mcgaugh98}; \cite{mortlock01};
see \cite{sanders02} for a recent review).  MOND postulates a
breakdown of Newton's law of gravity (or inertia) below few times
$a_0=1.2\times 10^{-8}$ cm s$^{-2}$ (\cite{begeman91}).  If this is
correct any physical system, not just galaxies, should deviate from
Newton's law below this acceleration. Confirming such a breakdown
would be of major importance regardless of the validity of MOND.

All these considerations prompted us to look for an experiment
involving gravitational acceleration of the order of $a_0$.  To reach
such low accelerations in a laboratory is clearly difficult, so we
focused our attention on a class of astrophysical objects useful for
this purpose.  Globular clusters (GC) are among the smallest
virialized structures in the universe and it is currently believed
that DM is not affecting their internal dynamics. This belief is
based on solid observational results:

1) the existence of tidal tails (\cite{leon00}) provides the strongest
evidence against large amounts of DM, either baryonic or non-baryonic,
in GCs. If present, the potential well would be so deep that no stars
could escape the cluster.  Numerical simulations (\cite{moore96}) show
indeed that in order to form tidal tails the {\it maximum}
mass-to-light ratio consistent with observations is $M/L < 2.5$ in
solar units.

2) In GC studies good agreement is found between dynamical and
luminous masses for an old stellar population (\cite{mandushev91}).

3) Up to now micro-lensing studies have failed to detect the dense
dark objects of GC dimension postulated by cold dark matter
hierarchical model and predicted to exist in the Milky Way halo
(\cite{navarro97}; \cite{ibata02}).

It is a fortunate coincidence that stars located in the outskirts of
GCs are subjected to a gravitational acceleration comparable to the
one experienced by stars in the outskirts of galaxies. Therefore,
these stars are ideal test particles for probing Newton's law down to
very weak acceleration regimes.

If as suggested by MOND, Newton's law really breaks down in the low
acceleration limit and a different law controls the motion of stars,
then the deviation from expectations must follow a systematic pattern.
That is, GC dynamics must reproduce exactly the dynamics of galaxies
(provided the regime of acceleration is the same). In particular,
compact GCs should behave like high-surface-brightness galaxies, i.e.,
the velocity dispersion profile should be flat at large radii
(\cite{carollo95}; \cite{simie02}). Loose GCs, on the other hand,
should behave like low-surface-brightness galaxies and show large mass
discrepancies at any radius (\cite{mateo98}).

In this letter, we present the result of a pilot experiment reaching
accelerations as low as $10^{-8}$ cm s$^{-2}$.

\begin{figure}
\centering
\includegraphics[height=8cm]{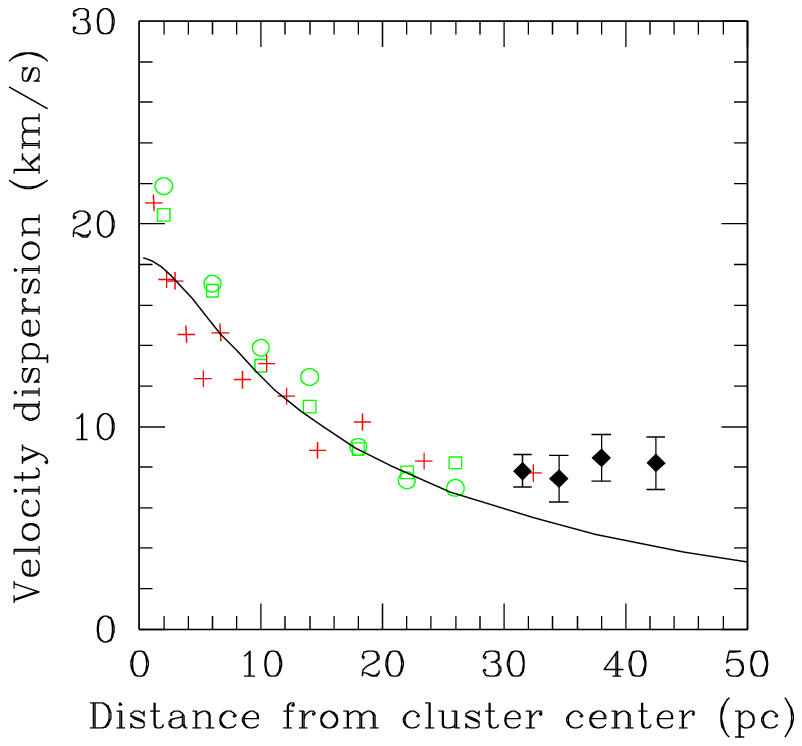}
\caption{\label{fig1} The velocity dispersion profile of $\omega$ Cen.
Proper motion data ({\bf Circles} and {\bf squares}) are from
\cite{vanleeuwen00}.  Radial velocities {\bf Crosses} are from
\cite{meylan95} up to 20 pc, while the last two points are from
\cite{meylan86}.  The solid line is the best fit model of radial
velocity data as derived by \cite{meylan95} (their fig 1).  Note the
good agreement between the last point of \cite{meylan86} and our
radial velocities measurement ({\bf diamonds}), which extend the
profile to $\sim 45$ pc.  }
\end{figure}

\section{A Pilot Experiment on $\omega$ Centauri}

We studied the dynamical properties of stars located in the outskirt
of $\omega$ Cen, a nearby, well studied GC, chosen only for being
one of the very few clusters for which internal proper motions are
known, and because it is visible from the south.

Located at a distance of 6.4 kpc from the Galaxy center, $\omega$ Cen
is the largest GC known, with mass estimates $>10^6$ M$_{\odot}$, and
tidal radius of $\sim 70$ pc (\cite{meylan86}).  In spite of the large
dimension and relative proximity to the Galaxy center, the density of
the Galactic halo (\cite{bahcall80}) is so low that within its tidal
radius $\sim 2\times 10^4$ M$_{\odot}$ of DM are expected, which is
dynamically negligible. The cluster is therefore suitable for testing
Newton's law down to the smallest acceleration probed by the motion of
its stars.

Available data allow the determination of the dynamical properties of
the cluster central regions. In Fig. 1 the radial velocity dispersion
profile, as derived using data for $\sim 400$ stars up to $\sim 30$ pc
from the center (\cite{meylan86}), is compared to the proper motion
dispersion profile derived using data for many thousands stars up to
25 pc from the cluster center (\cite{vanleeuwen00}).  The three
components of the velocity dispersion agree well with each other, with
no indication of anisotropy.  This is very important for two reasons:
i) it implies the cluster has reached dynamical equilibrium, making it
irrelevant whether the cluster is the result of merging; ii) radial
velocities, which are the easiest to determine and the simplest to
interpret, are representative of the cluster's kinematics.

To extend the dispersion profile and reach the desired acceleration
regime, we have selected from the list of van Leeuwen et al. 91 stars
at distance $>30$ pc from the center and membership probability
$>90$\%.  Selected stars are located only on the west side of the
cluster, covering $\sim 150$ degrees in azimuth.  This limitation is
due to the technical details of val Leeuwen et al. observations and is
in no way related to the physical properties of $\omega$ Cen. Thus
this should not bias our result toward large velocity dispersions.
 
Stars were subsequently observed during August 2001 with the ESO Very
Large Telescope (VLT), in order to measure their radial velocity.
Spectra were obtained with UVES (UV-visual Echelle Spectrograph) with
resolution R=40000.  The derived radial velocities have an average
error of 1 km s$^{-1}$, as verified using telluric lines.  With these
new data, one can in principle study the velocity dispersion of all 3
components of the velocity. However, due to the low velocity of stars
in the outskirt of the cluster, proper motion data have average errors
comparable to the dispersion we are trying to measure ($\sim 8$ km
s$^{-1}$). Therefore, though useful for selecting probable cluster
members, these data are not suitable for our purpose and we considered
radial velocity data only.  Having verified the isotropy of the
velocity ellipsoid, this limitation does not affect the generality of
the result.

\begin{table}[b]
\caption{Radial velocity dispersion}
\begin{tabular}{ccccc}
Bin   & N stars & $\sigma_{obs}$ & Rotation & $\sigma$\\
\hline
$30-33$ & 31 & 6.9 & 3.6 & 7.8$\pm$0.8\\
$33-36$ & 18 & 6.6 & 3.4 & 7.4$\pm$1.2\\
$36-40$ & 16 & 8.0 & 2.6 & 8.4$\pm$1.2\\
$40-45$ & 10 & 7.9 & 1.9 & 8.2$\pm$1.3\\
\end{tabular}
\end{table}

\section{Results for $\omega$ Centauri}

The average radial velocity of $\omega$ Cen is 232 km s$^{-1}$
(\cite{meylan86}), allowing the discrimination of cluster members from
foreground stars (Fig. 2).  Of the initial candidates, 75 were found to be 
true cluster members (consistent with the assigned membership
probability). The remaining have radial velocity of few km s$^{-1}$
and are most likely part of the Galactic disk.

The radial velocity dispersion as a function of distance is given in
Table 1. Columns 1 and 2 give the bin width in pc and the number of
stars per bin. Column 3 gives the observed velocity dispersion in km
s$^{-1}$.  In order to trace the cluster potential, these values were
combined in quadrature with the values for the cluster rotation given
in Column 4 (\cite{meylan86}).  This correction is marginal and is
relevant only for comparing our data with published data.  The final
corrected values of the velocity dispersion $\sigma$ are in Column 5
These new data extend the study of the kinematics from 30 to 45 pc
from the cluster center (projected distances, Fig. 1). We found the
dispersion does not decrease monotonically with distance, as expected
in Newtonian dynamics, remaining constant at large radii.  Invoking a
rapid increase of the velocity anisotropy outward of 25 pc from the
center to make the profile falling as $1/\sqrt{r}$, implies that the
other two components of the velocity dispersion have only $\sim 70$\%
of the energy of the radial component. Such a large and rapid increase
of the velocity anisotropy seems very unlikely.  Therefore, though
limited to the west side of the cluster, the apparent constancy of the
velocity dispersion is intriguing.

\begin{figure}
\centering
\includegraphics[height=5cm]{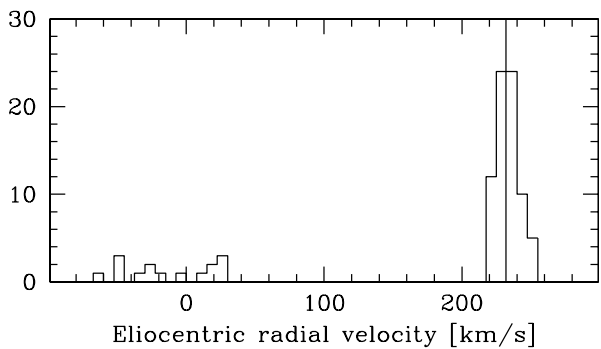}
\caption{\label{fig2} 
Distribution of the heliocentric radial velocity
for the 91 selected stars.  Of all targets, 75 are clearly members
having radial velocity comparable to the one of the cluster, marked by a
vertical line. The remaining ones are most probably part of the
Galactic disk.}
\end{figure}

\section{Discussion}

This project was started as a pilot experiment to see if at least in
one globular cluster interesting results, either in favor or against Newton's
law of gravity, could be found. After the study of $\omega$ Cen was
completed, we became aware of the existence of a similar set of data
showing that also in M15 the velocity dispersion profile remains flat
(within errors) at large radii (see Fig. 8 in \cite{drukier98}).

The flattening of the velocity dispersion profile at large radii
observed in both $\omega$ Cen and M15 can be due i) to tidal heating,
as explicitly suggested for M15 (\cite{drukier98}); ii) to a DM halo
surrounding the clusters; iii) to a breakdown of Newton's law in
the weak regime limit.
 
The first two scenarios are certainly possible. Despite the fact that
they require ad hoc assumptions to maintain the profile flat, they can
not be ruled out by the present data. Therefore, the purpose of this
work is not to disprove these two scenarios, but to fully investigate
the third one.

In an attempt to study Newton's law in a regime were its validity is
not proved, the use of masses (required to compute accelerations)
derived from dynamical models based on this law would lead to a
circular argument.  In particular masses derived from the virial
theorem or King models may be overestimated (one should keep in mind
that both man-made spacecrafts and MOND suggest gravity is pulling
harder than expected).  To estimate the mass of the clusters we
therefore are better off converting the cluster luminosities into
masses adopting a suitable mass-to-light ratio $\tau$.

The total absolute V magnitude of $\omega$ Cen and M15 are $-10.29$
and $-9.17$ (\cite{harris96}), corresponding to a mass $M=1.08\tau$
and $3.87\tau$ millions of solar masses, respectively.  The mass so
estimated can be safely assumed to be all inside the radius at which
the velocity dispersion profile becomes flat.  In the case of $\omega$
Cen the profile flattens at $27\pm 3$ pc from the center, where the
acceleration is $2.1^{+0.5}_{-0.4} \times 10^{-8} \times \tau$ cm
s$^{-2}$. In M15 the profile flattens at $18 \pm 3$ pc (that is $6\pm
1$ arcmin from the center, see Fig. 8 of \cite{drukier98}),
corresponding to $1.7^{+0.7}_{-0.5}\times 10^{-8} \times \tau$ cm
s$^{-2}$.  Very interestingly, provided $\tau$ in the same in the two
clusters, both profiles start to be flat for the same value (within
errors) of the acceleration.

These two clusters have very little in common: they have different
masses, different positions and orbits in the galactic halo.  Their
dynamical evolution was also different.  Moreover, it has been claimed
that $\omega$ Cen could contain two stellar populations being the
result of a merging of two cluster (\cite{lee99}), or that it is the
remnant of a dwarf galaxy (\cite{hilker00}; \cite{carraro00}).  Thus we
find the fact that the two profiles are so similar a significant one.

To further investigate if this result is in agreement with the claim
that deviations from Newtonian dynamics should appear at few times
$a_0$ (\cite{milgrom83}), we need to explicitly adopt a value for
$\tau$.  For a Salpeter's initial mass function with slope $x=1.3$,
updated theoretical evolutionary models (\cite{cassisi98}) give
$M/L=1.4$ for simple stellar population.  This value would apply to a
GC that has retained all its initial mass. Real clusters do experience
mass losses because of tidal interaction and evaporation
(e.g. \cite{aguilar88}, \cite{smith02}).  The dynamical evolution
preferentially removes low mass stars that are the major contributors
to the cluster mass, while contributing little to the luminosity.
This effect has been verified in those GCs where the present day mass
function has been determined, finding a slope $x=0.7$ at low mass end
(\cite{demarchi99}; \cite{piotto97}; \cite{andreuzzi00}).  With this
mass function the same models give $M/L=0.6$. Considering that both
values depend on the adopted low mass cutoff, we assume $M/L=1$ as a
fair estimate of the true value. For this $\tau$ the velocity
dispersion profile of both clusters starts to be flat at $\sim 2 a_0$,
consistent with MOND prediction.

GCs are hundreds of times smaller and thousands of time lighter than
galaxies, nonetheless the velocity dispersion profile of at least
these two clusters precisely mimics, both in shape and absolute
acceleration, the one of elliptical galaxies (explained invoking DM;
\cite{carollo95}).  There is no reason for the flattening in GCs and
galaxies to occur for the same value of the acceleration.  If tidal
heating is responsible for the observed profile, then we should
conclude that the tidal action of the Milky Way not only conspires to
make $\omega$ Cen and M15 very similar to each other, but also
produces exactly the same effect as observed in galaxies.  The same
argument applies if a DM halo surrounding these two clusters is
invoked to explain the effect.  As we said, both scenarios are
possible but  require {\it ad hoc} assumptions  to explain 
these coincidences, making them somewhat weak.

The parallel between GCs and galaxies can be pushed even further if we
reanalyze a recent result for the GC Palomar 13.  It has been reported
that this GC has a velocity dispersion of 2 km/s, corresponding to
$M/L \sim 40$ (\cite{cote02}). This high dispersion can simply
indicate that the cluster is out of dynamical equilibrium, in which
case the virial theorem is not applicable and the derived M/L
meaningless.  Palomar 13, however, moves in an hardly disturbed orbit
(\cite{siegel01}) and its crossing time ($\sim 2$ My for an internal
dispersion of 2 km s$^{-1}$) is short enough to ensure stars reach
dynamical equilibrium quickly. We therefore argue there are good
reasons to believe Palomar 13 is in equilibrium. If this is the case,
the data are either reflecting the effects of a massive DM halo or a
break down of Newton's law.  Palomar 13 is very loose, having central
density of only $\sim 4$ M$_{\odot}$ pc$^{-3}$ (\cite{harris96}),
corresponding to an internal acceleration of gravity $<10^{-9}$ cm
s$^{-2}$ all the way to the cluster center. It is therefore similar to
low-surface-brightness galaxies which are characterized by very large
$M/L$ (\cite{mateo98}).  If our interpretation of the result for
$\omega$ Cen and M15 is correct, it is therefore not surprising that
Palomar 13 shows a large mass discrepancy because this is expected in
the case of a break down of Newton's law below $\sim a_0$. Indeed, it
as been shown that both Palomar 13 \cite{scarpa03} and
low-surface-brightness galaxies \cite{mcgaugh98} follows the
predictions of MOND.

\section{Conclusions}

We have presented new measurements of the velocity dispersion of the
GC $\omega$ Cen, which allow us to trace the gravitational potential
down to an acceleration of $8\times 10^{-9}$ cm s$^{-2}$.  We have
found that the dispersion profile remains flat well inside the tidal
radius as soon as the acceleration of gravity approaches $a_0$. A
similar behavior is also observed in M15, though it was previously
ascribed to tidal heating (\cite{drukier98}).

This result is surprising and may suggest a failure of Newton's law
at low accelerations. A conclusion also supported by data for Palomar
13.

Although present data can not rule out the standard dark-matter
scenario, nor tidal heating effects, the coincidence we pointed out is
remarkable.  Combined with the anomalous acceleration experienced by
spacecrafts in the periphery of the solar system (\cite{anderson98}),
this result prompts for accurate determination of the internal
dynamics of as many GCs as possible. Beside the interest of testing a
law of physics in a regime where its validity has never been verified,
these observations have the potential of proving or disproving in a
direct way the existence of non-baryonic DM.

We wish to thanks L. Pulone and L. Ciotti and S. Cassisi for helpful
discussion and suggestion.

\end{document}